\documentclass{ws-ijmpa}
\usepackage[super,compress]{cite}
\usepackage{graphicx}
\usepackage{braket}
\usepackage{bm}
\usepackage{comment}
\usepackage{here}
\usepackage{multirow}
\usepackage{color}

\def\Journal#1#2#3#4{{#1} {\bf #2}, #3 (#4)}
 	
\def\ARNPS{Annu. Rev. Nucl. Part. Sci.} 
\def\AandA{Astron. Astrophys.} 

\def\APJ{Astrophys. J.}
\def\APJL{Astrophys. J. Lett.}

\def\CMP{Commn. Math. Phys.}
\def\EPJC{Eur. Phys. J. C}

\def\JCAP{J. Cosmol. Astropart. Phys.}

\def\MNRAS{Mon. Not. R. Astron. Soc.}
\def\MPLA{Mod. Phys. Lett. A}

\def\NATU{Nature}

\def\PL{Phys. Lett.}
\def\PLB{{Phys. Lett.} B}

\def\PPNP{Prog. Part. Nucl. Phys.}

\def\PRL{Phys. Rev. Lett.}
\def\PRA{Phys. Rev. A}
\def\PRD{Phys. Rev. D}

\def\PRR{Phys. Rev. Res.}

\def\PTEP{Prog. Theor. Exp. Phys.}

\def\RPP{Rep. Prog. Phys.}

\begin{document}
\markboth{A. Takeshita and T. Kitabayashi}{Interstellar medium gas heating by primordial black holes and dark matter particles}

%
\catchline{}{}{}{}{}
%


\title{Interstellar medium gas heating by primordial black holes and dark matter particles}

\author{Amane Takeshita}

\address{Graduate School of Science, Tokai University,\\
4-1-1 Kitakaname, Hiratsuka, Kanagawa 259-1292, Japan}

\author{Teruyuki Kitabayashi\footnote{Corresponding author}}

\address{Department of Physics, Tokai University,\\
4-1-1 Kitakaname, Hiratsuka, Kanagawa 259-1292, Japan\\
teruyuki@tokai.ac.jp}

\maketitle

\begin{history}
\received{Day Month Year}
\revised{Day Month Year}
\end{history}

\begin{abstract}
The Leo T dwarf galaxy has been utilized to investigate the heating of interstellar medium gas by both primordial black holes (PBHs) and dark matter (DM) particles. Previous studies have typically assumed that either PBHs or DM particles are responsible for heating the interstellar medium gas. In contrast, this study considers the simultaneous contribution of both PBHs and DM particles to the heating process. 

If both PBHs and dark photons heat the gas in Leo T, a stringent constraint on the PBH fraction, $f_{\rm PBH}=\rho_{\rm PBH}/\rho_{\rm DM}$ is obtained for $4 \lesssim M_{\rm PBH}/M_\odot \lesssim 10^2$, where $\rho_{\rm PBH}$, $\rho_{\rm DM}$,  represent the energy densities of PBHs and DM, respectively, and $M_{\rm PBH}$, and $M_\odot$ denote the masses of PBHs and Sun, respectively. Conversely, if both PBHs and millicharged particles heat the gas, it becomes challenging to impose a more significant constraint on the PBH fraction than previously achieved, due to the very small allowed values of charge parameters within the model.
\end{abstract}

\ccode{PACS numbers:14.60.Pq}


\section{Introduction\label{section:introduction}}
Dark matter (DM), the enigmatic source of gravity in the universe, remains one of the fundamental challenges in particle physics and cosmology \cite{Arbey2021PPNP}. Among the most compelling candidates for DM is the primordial black hole (PBH) \cite{Carr1975APJ,Carr2020ARNPS,Carr2021RPP,Auffinger2023PPNP,Khlopov2010RAA,Belotsky2014MPLA,Belotsky2019EPJC,Heydari2022EPJC,Heydari2022JCAP,Heydari2024JCAP,Heydari2024EPJC,Heydari2024ApJ}, which formed during the early universe. While PBHs lose mass through the emission of particles via Hawking radiation \cite{Hawking1975CMP}, those with an initial mass exceeding $10^{15}$ g can survive in the current universe without fully evaporating. Although  it is unlikely that all DM is composed solely of PBHs across the broad mass spectrum \cite{Carr2021RPP}, the possibility of PBH formation in numerous early universe models suggest it is reasonable to consider at least a portion of DM consists of PBHs.

If only a portion of DM consists of PBHs, what constitutes the remainder? Since DM has eluded detection through electromagnetic observations, it is typically assumed to lack electromagnetic interactions. Under this assumption, electrically neutral, massive, and stable particles predicted by theories beyond the standard model are considered promising DM candidates  \cite{Arbey2021PPNP}. However, some particle DM models propose weak electromagnetic interactions. For example, dark photons may couple very weakly to electrically charged particles via kinetic mixing with ordinary photon \cite{Dubovsky2015JCAP}. If dark photons is massive, they could serve as DM. Millicharged particles are another potential DM candidate \cite{Holdom1986PLB}. Additionally, the  temperature  at cosmic dawn being only half of the expected value \cite{Bowman2018Nature} can be explained by a Coulomb-like interaction between millicharged particles and neutral hydrogen (${\rm H_I}$) \cite{Barkana2018Nature}.

In recent years, the Leo T dwarf galaxy \cite{IrwinAPJ2007}, along with other astrophysical gas clouds, has been used to study the heating of interstellar medium gas by PBHs and DM particles. Leo T, located in the Leo constellation, is classified as a transitional object between dwarf spheroidal and dwarf irregular galaxies. Unlike dwarf galaxies within the Milky Way's virial radius, which often lack gas due to ram pressure stripping \cite{Gunn1972APJ,Grcevich2010APJ}, Leo T, being beyond this radius, can retain significant amounts of gas. When DMs (such as PBHs, dark photons, or millicharged particles) heat the interstellar medium, the heating rate $Q$ must be equal or less than the astrophysical cooling rate $C$, $Q \le C$. If $Q$ exceeds $C$, the temperature of the interstellar medium gas would increase, potentially altering the ionization ratio of the gas \cite{Kim2021MNRAS,Lu2021APJL,Laha2021PLB,Wadekar2021PRD,Takhistov2022JCAP,Wadekar2023PRD,Melikhov2023PRD,Dubovsky2015JCAP}. In the literature, the heating rate and the cooling rate are depicted as $\dot{Q}$ and $\dot{C}$, respectively; however, we omit `dot' in this paper. 

Previously, various sources have been considered for heating interstellar medium gas, including  PBHs via Hawking radiation \cite{Kim2021MNRAS,Laha2021PLB,Melikhov2023PRD}, PBHs via accretion disks \cite{Lu2021APJL,Takhistov2022JCAP,Wadekar2023PRD}, compact DM  \cite{Wadekar2023PRD}, dark photons (hidden photons) \cite{Dubovsky2015JCAP,Bhoonah2019PRD,Bhoonah2021PRD,Wadekar2021PRD,Prabhu2023PRD}, millicharged particles \cite{Bhoonah2018PRL,Wadekar2021PRD}, vector portal DM \cite{Bhoonah2019PRD}, Higgs portal DM \cite{Wadekar2022PRD}, sterile neutrinos \cite{Wadekar2022PRD}, dark baryons \cite{Wadekar2022PRD}, axions and axion like particles \cite{Wadekar2023PRD, Wadekar2022PRD}, excited DM state \cite{Wadekar2022PRD}, magnetically charged DM \cite{Wadekar2023PRD}, composite DM \cite{Bhoonah2021PRD,Wadekar2023PRD}, annihilating DM \cite{Chen2023MNRAS}, electromagnetic dipole moment of DM \cite{Bi2023EPJC}, and DM-baryon interactions \cite{Bhoonah2019PRD, Shoji2024MNRAS}. These studies assume that only one type of DM is responsible for heating the interstellar medium gas. For example, in Refs \cite{Kim2021MNRAS,Laha2021PLB,Melikhov2023PRD,Lu2021APJL,Takhistov2022JCAP}, PBHs are considered to be part of the DM and responsible for heating the gas, while other DM are not. Similarly, Ref. \cite{Bhoonah2018PRL} assumes that millicharged particles are the sole contributors to the heating of the gas, with other DM having no effect.

In this study, we consider DM to consist of two components: PBHs and additional DM particles. Unlike previous studies \cite{Kim2021MNRAS,Laha2021PLB,Melikhov2023PRD,Lu2021APJL,Takhistov2022JCAP}, which assume that only one type of DM particles heats the gas, we propose that both PBHs and DM particles contribute to the heating of the gas simultaneously in the Leo T dwarf galaxy. We specifically consider dark photons or millichaged particles as the DM particles \textcolor{black}{\footnote{\textcolor{black}{The axions and axion like particles are also good candidate of DM in modern situation \cite{Wadekar2023PRD, Wadekar2022PRD}. The gas heating via axions and PBHs simultaneously should be investigate. We would like to intend to study this issue in more detail in the future.}}}. The heating mechanism for PBHs is either through Hawking radiation or heat transport from the accretion disk. In contrast, the DM particles heat the gas via mechanism specific to their particle models. 

The remainder of this article is organized as follows. Section \ref{section:GasHeating} provides a brief review of the Leo T profile and the current constraints on PBHs. Before exploring the simultaneous heating of the gas in Leo T by both PBHs and DM particles, we first estimate the  heating contribution from individual sources, including light PBHs, heavy PBHs, dark photons, and millichaged particles. The main discussion in section \ref{section:GasHeating} are based on Ref. \cite{Laha2021PLB} for light PBHs, Refs.\cite{Lu2021APJL,Takhistov2022JCAP} for heavy PBHs, and Ref. \cite{Wadekar2021PRD} for dark photons and millicharged particles. In section \ref{section:PBH_and_DM}, we assess the combined effect of PBHs and DM particles (dark photons or millicharged particles) on gas heating. We discuss constraints on the PBH fraction, the kinetic mixing parameter of dark photons, and the charge parameter for millicharged particles. The article concludes with a summary of the key findings in section \ref{section:summary}. 

In this paper, we use the natural units where $c=\hbar=k_{\rm B} = 1$. Here, $c$, $\hbar$ and $k_{\rm B}$ denotes the speed of light, $\hbar = h/(2\pi)$ represents the reduced Planck constant, and the Boltzmann constant.

\section{Gas heating by individual source \label{section:GasHeating}}
\subsection{Leo T profile \label{section:LeoT_profile}}

\textcolor{black}{The physical parameters of Leo T are derived from the measured density and velocity dispersion of ${\rm H_I}$ gas \cite{Faerman2013APJ, Ryan-Weber2008MNRAS} using an appropriate model. In this study, we adopt the model of Leo T in Ref. \cite{Faerman2013APJ}.}

The gas in the inner region of Leo T (region I) is predominantly atomic hydrogen, while the gas outside is highly ionized \cite{Faerman2013APJ}. Given that free electrons in the ionized region cool very efficiently \cite{Smith2017MNRAS}, we focus on the gas heating in region I. The radius of region I is $r_{\rm I} \simeq 0.35$ kpc. In region I, the average ${\rm H_I}$ gas density is $n_{\rm H} = 0.07$ ${\rm cm^{-3}}$, electron density is $n_e = 0.001$  ${\rm cm^{-3}}$, and the DM energy density is $\rho_{\rm DM} = 1.75$ GeV ${\rm cm^{-3}}$ \cite{Simon2007APJ,Faerman2013APJ,Wadekar2021PRD,Laha2021PLB}. The gas column density is $m_{\rm H} n_{\rm H} r_{\rm sys} = 1.265 \times 10^{-4}$  g ${\rm cm}^{-2}$, where $m_{\rm H}$ is the mass of neutral hydrogen and $r_{\rm sys} = r_{\rm I}$ is the size of the region. The velocity dispersion of the neutral hydrogen (${\rm H_I}$) gas is approximately $\sigma_g = 6.9$ km/s \cite{Ryan-Weber2008MNRAS,Simon2007APJ,Adams2018AA}, leading to an estimated gas temperature of  around $T=6000$ K in the region I.

We use the following approximate formula to calculate the gas cooling rate \cite{Laha2021PLB,Wadekar2023PRD,Smith2017MNRAS}
\begin{equation}
C=n_{\rm H}^2 10^{[{\rm Fe}]/[{\rm H}]} \Lambda (T),
\end{equation}
where $[{\rm Fe}]/[{\rm H}]=\log_{10}(n_{\rm Fe}) / (n_{\rm H})_{\rm gas}$ - $\log_{10}(n_{\rm Fe}) / (n_{\rm H})_{\rm Sun}$ = $-1.74$  represents the metallicity relative to the Sun\textcolor{black}{\cite{Kirby2013APJ}}, and $\Lambda(T) = 2.51 \times 10^{-28} T^{0.6}$ is the cooling function valid  for $300 ~ {\rm K} < T < 8000~ {\rm K}$ \cite{Wadekar2021PRD}.  For this study, we adopt the value
\begin{equation}
C=2.28\times 10^{-30} ~ {\rm erg} ~ {\rm cm}^{-3} ~ {\rm s}^{-1},
\end{equation}
as cited in \textcolor{black}{Refs.} \cite{Lu2021APJL,Laha2021PLB}. \textcolor{black}{\footnote{\textcolor{black}{In the  region I, the ${\rm H_I}$ density is approximately constant \cite{Wadekar2021PRD}. For this reason, the gas temperature and the cooling rate are assumed to be constant in this study.}}}

We will use $n_{\rm He}/n_{\rm H}=0.08$ to calculate the interactions of millicharged particles with the interstellar medium gas \textcolor{black}{\cite{Kirby2013APJ,Wadekar2021PRD}}.

\subsection{PBH constraints \label{section:PBH_constraint}}
The abundance of PBH is typically expressed as the ratio
\begin{equation}
f_{\rm PBH} = \frac{\rho_{\rm PBH}}{\rho_{\rm DM}} (<1), 
\end{equation}
where $\rho_{\rm PBH}$ represents the energy density of PBHs. Current constraints on this ratio are approximately (see FIG.10 in Ref.\cite{Carr2021RPP} and FIG.2 in Ref. \cite{Laha2021PLB})
\begin{equation}
\begin{cases}
f_{\rm PBH} \lesssim 4 \times 10^{-4} &   (M_{\rm PBH}{\rm [g]} \lesssim 1\times 10^{16})  \\
4 \times 10^{-4} \lesssim f_{\rm PBH} \lesssim 0.05 & (1 \times 10^{16} \lesssim M_{\rm PBH}{\rm [g]}  \lesssim 4\times 10^{16})\\
0.05 \lesssim f_{\rm PBH} \lesssim 1 & (4\times 10^{16} \lesssim M_{\rm PBH}{\rm [g]} \lesssim 1\times 10^{17}),
\end{cases}
\label{Eq:lightPBHmass}
\end{equation}
for $M_{\rm PBH} \lesssim 10^{17}$ g, where $M_{\rm PBH}$ denotes the initial mass of PBH. \textcolor{black}{The first line in Eq. (\ref{Eq:lightPBHmass}) seems to have to be dependent from mass of PBH; however, it is just an approximation.} In this paper, the PBHs with $M_{\rm PBH} \lesssim 10^{17}$ g are referred to as the light PBHs. It is known that all DMs could be composed of light PBHs only if $M_{\rm PBH}  \simeq 1\times 10^{17} ~ {\rm g}$ \cite{Carr2021RPP,Laha2021PLB}. In other mass ranges, light PBHs would constitute only a small fraction of DM \cite{Carr2021RPP,Laha2021PLB}. For light PBHs, no accretion disk forms, and only Hawking radiation is available for heating the gas \cite{Kim2021MNRAS,Laha2021PLB,Melikhov2023PRD}. 

In contrast, for PBH with $M_{\rm PBH}/M_\odot \gtrsim 0.1$, we classify these as heavy PBH, where $M_\odot = 1.99\times 10^{33}$ g is the solar mass. For such PBHs, Hawking radiation becomes negligible and heat transport from the accretion disk is the primary mechanism for gas heating \cite{Lu2021APJL,Takhistov2022JCAP}. The allowed ratio of $f_{\rm PBH}$ for heavy PBHs is approximately (see FIG.10 in Ref.\cite{Carr2021RPP} and FIG.5 in Ref.\cite{Takhistov2022JCAP}) 
\begin{equation}
\begin{cases}
f_{\rm PBH} \simeq 0.08 & (0.1 \lesssim M_{\rm PBH}/M_\odot \lesssim 0.2)  \\
0.02 \lesssim f_{\rm PBH} \lesssim 0.08 & (0.2 \lesssim M_{\rm PBH}/M_\odot \lesssim 1)  \\
f_{\rm PBH} \simeq 0.08 & (1 \lesssim M_{\rm PBH}/M_\odot \lesssim 30)  \\
10^{-8} \lesssim f_{\rm PBH} \lesssim 10^{-3} & (30 \lesssim M_{\rm PBH}/M_\odot \lesssim 3\times 10^4)  \\
f_{\rm PBH} = {\rm almost ~vanish} & (3\times 10^4 \lesssim M_{\rm PBH}/M_\odot).
\end{cases}
\end{equation}

For PBH with intermediate masses, \textcolor{black}{$10^{17} {\rm g} \lesssim M_{\rm PBH} \lesssim 0.1 M_\odot$}, neither Hawking radiation nor heat transport from the accretion disk significantly impacts gas heating. 

In this study, we primary focus on light PBHs with masses ranging from $4\times 10^{16} ~ {\rm g} \lesssim M_{\rm PBH} \lesssim 1\times 10^{17}~ {\rm g} $ and heavy PBHs masses between $0.1 \lesssim M_{\rm PBH}/M_\odot \lesssim 30$.

\subsection{Gas heating by light PBHs \label{section:PBH_light}}
Before examining scenarios where both PBHs and DM particles contribute to heating the interstellar medium gas in Leo T, we first estimate the gas heating ratio from individual heating sources. In this subsection, we consider light PBHs as the sole heating source for the gas in Leo T. 

For light PBHs, no accretion disk forms, and thus only Hawking radiation contributes to gas heating \cite{Kim2021MNRAS,Laha2021PLB,Melikhov2023PRD}. The heating rate per unit time due to particle $i=(\gamma, e^\pm)$ emission via Hawking radiation from a single PBH is given by \cite{Laha2021PLB}
\begin{align}
H_{i{\rm PBH}} = & \int_0^\infty f_h(E)(E-m_i) \frac{d^2N_i(M_{\rm PBH}, a_\ast)}{dtdE}  \left( 1-e^{-\tau_i} \right) dE,
\end{align}
where $m_i$ is the mass of particle $i$, $a_* = J_{\rm PBH}/(GM_{\rm PBH}^2)$ is the reduced spin Kerr parameter (with $G$ being the gravitational constant and $J_{\rm PBH}$ the angular momentum of the PBH), and $f_h(E) = 0.367+0.395(11 {\rm eV}/(E-m_i))^{0.7}$ represents the fraction of energy loss deposited as heat \cite{Kim2021MNRAS,Laha2021PLB,Shull1985APJ,Ricotti2002APJ,Furlanetto2010MNRAS}.  \textcolor{black}{(See also Refs.\cite{Slatyer2013PRD,Madhavacheril2014PRD,Slatyer2016PRD1,Slatyer2016PRD2,Liu2016PRD,Slatyer2017PRD,Clark2017PRD})}. The PBH emission spectrum, denotes the number of particles  $N_i$ emitted per unit energy per unit time, is given by
\begin{equation}
\frac{d^2N_i}{dtdE} = \frac{1}{2\pi}\sum _{\rm dof}\frac{\Gamma_i(E, M_{\rm PBH}, a_*)}{e^{E'/T_{\rm PBH}} \pm 1},
\end{equation}
where $\Gamma$ represents the graybody factor\textcolor{black}{\cite{MacGibbon1990PRD}} and $E'$ denotes the total energy of a particle around rotating PBH. The sign `+' applies to Fermions, while `-' applies to Bosons. The temperature of a charge-neutral rotating PBH is expressed as
\begin{equation}
T_{\rm PBH} = \frac{1}{4\pi G M_{\rm PBH}}\left(\frac{\sqrt{1-a_*^2}}{1+\sqrt{1-a_*^2} } \right).
\end{equation}
In this study, the PBH emission spectra for $\gamma, e^\pm$ are generated numerically using the BlackHawk code \cite{Arbey2019EPJC}. The optical depths of gas for $\gamma$ and $e^\pm$ are given by
\begin{eqnarray}
 \tau_\gamma = \frac{m_{\rm H} n_{\rm  H} r_{\rm sys}}{\lambda},  \quad \tau_{e^\pm} =  \frac{m_{\rm H} n_{\rm  H} r_{\rm sys} S(E)}{E},
\end{eqnarray}
where $\lambda$ and $S(E)$ represent the absorption length \textcolor{black}{in unit of $\rm{g~cm^{-2}}$} and the stopping power \textcolor{black}{in unit of $\rm{MeV ~(g ~ cm^{-2})^{-1}}$}, respectively. We utilize the cumulative photon absorption length from Ref. \cite{PDG} and the electron stopping power in hydrogen gas from the NIST database \cite{Berger2007}. 

The heating rate per unit volume per unit time due to the emission of particle $i$ from PBHs is given by
\begin{equation}
Q_{i{\rm PBH}} = n_{\rm PBH} H_{i{\rm PBH}}, 
\end{equation}
where 
\begin{equation}
n_{\rm PBH} = \frac{f_{\rm PBH}  \rho_{\rm DM}}{M_{\rm PBH}}, 
\end{equation}
denotes the number density of PBH. The total heating rate per unit volume per unit time from photons, electrons, and positrons emitted by PBHs via Hawking radiation is given by
\begin{equation}
Q_{\rm PBH,HR} = Q_{\gamma{\rm PBH}} + Q_{e^-{\rm PBH}} + Q_{e^+{\rm PBH} }.
\end{equation}
%
\begin{figure}[t]
\centering
\includegraphics[keepaspectratio, scale=0.5]{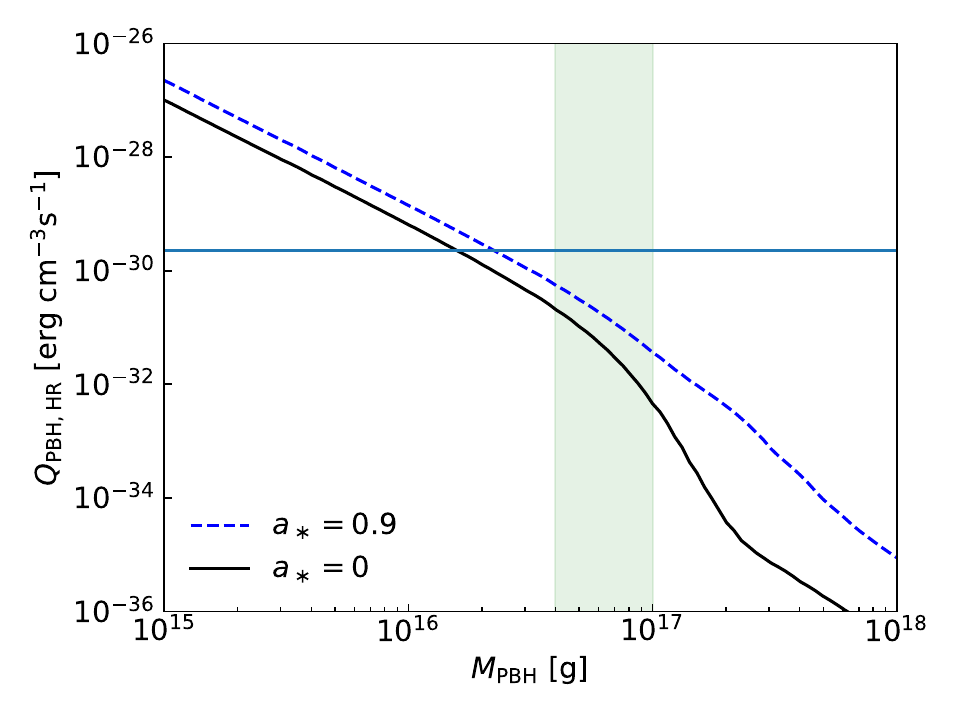}
 \caption{Gas heating rate in Leo T due to Hawking radiation from light PBHs. The curve illustrates the heating rate for the scenario where all DMs consists of light PBHs. The horizontal line represents the gas cooling rate in Leo T, $C=2.28\times 10^{-30} ~ {\rm erg} ~ {\rm cm}^{-3} ~ {\rm s}^{-1}$. }
 \label{FIG:pbh_Q_HR_gram}
 \end{figure}

FIG. \ref{FIG:pbh_Q_HR_gram} shows the gas heating rate in Leo T due to Hawking radiation from light PBHs. The curve illustrates the heating rate assuming all DMs are composed of light PBHs, representing the maximum gas heating rate for a specific case with $f_{\rm PBH} = 1$. For comparison, the horizontal line denotes the gas cooling rate in Leo T, $C=2.28\times 10^{-30} ~ {\rm erg} ~ {\rm cm}^{-3} ~ {\rm s}^{-1}$. 

Light PBHs could constitute all DMs in the very narrow mass range around $M_{\rm PBH}  \simeq 1\times 10^{17} ~ {\rm g}$. In other mass ranges, light PBHs can only represent a small fraction of the DM \cite{Carr2021RPP,Laha2021PLB}. Therefore, considering the current constraints of $f_{\rm PBH}$, the actual heating ratio in Leo T must be lower than what is shown in FIG. \ref{FIG:pbh_Q_HR_gram}. For example, with the constraint $0.05 \lesssim f_{\rm PBH} \lesssim 1$, only the mass region $4\times 10^{16} ~ {\rm g} \lesssim M_{\rm PBH}  \lesssim 1\times 10^{17} ~ {\rm g}$ in  green band in the FIG. \ref{FIG:pbh_Q_HR_gram} is relevant. Even using the overestimated curve in FIG. \ref{FIG:pbh_Q_HR_gram}, the ratio $Q/C$ falls within small as $1.8 \times 10^{-3} \lesssim Q/C \lesssim 8.5 \times 10^{-2} $ for $a_*=0$ (Schwarzschild BH) and $1.6 \times 10^{-2} \lesssim Q/C \lesssim 2.4 \times 10^{-1} $ for $a_*=0.9$ (rotating BH). 

Since $Q_{\rm PBH,HR} < C$ is always satisfied (in other words, $Q_{\rm PBH,HR}$ cannot be comparable to $C$) in this mass range, it is challenging stronger constraints on light PBHs than those previously claimed Ref. \cite{Laha2021PLB}. \textcolor{black}{We would like to note that the possible suppression of Hawking radiation related to the information paradox has recently been discovered \cite{Dvali2016FortschrPhys,Dvali2020PRD,Alexandre2024PRD,Dvali2024PRD}. The inclusion of such effects, e.g. memory burden effect, may change the results of this paper. We would like to keep this issue in our future study.}

\subsection{Gas heating by heavy PBHs \label{section:PBH_heavy}}
In this subsection, we consider that the heavy PBHs are the sole heating source for the gas in Leo T. For heavy PBHs, Hawking radiation is negligible, and heat transport from the accretion disk becomes the primary mechanism for gas heating \cite{Lu2021APJL,Takhistov2022JCAP}.  In this case, the three main mechanisms for gas heating are accretion photon emission, dynamical friction, and accretion mass outflow. We assume all heavy PBHs are Schwarzschild black holes and employ the methods and numerical values for model parameters as described in Model 2 of Ref.\cite{Takhistov2022JCAP}.  For the mass range $0.1 \le M_{\rm PBH}/M_\odot \le 30$ (including which is the primary focus of this study), accretion mass outflow is the dominant heating mechanism \cite{Takhistov2022JCAP}.

Accretion mass outflows, primarily composed protons, become significant in the advection-dominated accretion flow regime within the accretion disk. The heat deposited by these protons is limited by the electronic stopping cross section for protons passing through hydrogen, denoted as $S(E)$, which we obtain from FIG.9 in Ref.\cite{Bailey2019PRA}. The energy deposited per proton is given by
\begin{equation}
\Delta E = {\rm min} (E, n_{\rm H}r_{\rm sys}S(E)),
\end{equation}
where $n_{\rm H}r_{\rm sys}=7.56\times 10^{19}$ cm$^{-2}$ represents the gas column density in Leo T.

\textcolor{black}{The outflows reduce the falling gases and the accretion rate at smaller radii. The outflow rate can be approximated by a self-similar power-law \cite{Blandford1999MNRAS}
\begin{equation}
\dot{M}_{\rm out}(r) = \dot{M}_{\rm in}(r_{\rm out})\left( \frac{r}{r_{\rm out}} \right)^s,
\end{equation}
where $r_{\rm out}$ is outer radius and $\dot{M}_{\rm in}(r_{\rm out})$ is the Bondi-Hoyle-Lyttleton accretion rate. The energy deposit from streaming outflow of proton can be estimated by the convolution of the proton emission with the heat generated per proton $\Delta E$. According to the formalism in Ref. \cite{Lu2021APJL, Takhistov2022JCAP}, the heating rate is estimated as 
\begin{equation}
H_{\rm PBH, outflow} = \int_{r_{\rm in}}^{r_{\rm out}}  \frac{\mathcal{D} f_h \Delta E}{\mu m_p} \frac{d \dot{M}_{\rm out}}{dr}dr
=
\int_{E(r_{\rm in})}^{E(r_{\rm out})} f_h \Delta E f(E) dE,
\end{equation}
where, $\mathcal{D}$ denotes the duty cycle factor, $\mu$ denotes the mean molecular weight.  The fraction of energy deposited as heat $f_h$ is taken to be $1/3$. } The inner disk radius is taken to be the innermost stable circular orbit (ISCO) of a test particle, $r_{\rm in} = r_{\rm ISCO}=3r_{\rm s}$, where $r_{\rm s} = 2G\textcolor{black}{M_{\rm PBH}}$ is the Schwarzschild radius. The outer radius is set to $r_{\rm out} = 100 r_{\rm s}$. The kinetic energy of a proton ($m_p = 1.67 \times 10^{-27}$ kg = 938.2 MeV) at a radius $r$ is determined based on the Keplerian velocity, \textcolor{black}{$v(r) \propto 1/\sqrt{r}$}, as
\begin{equation}
E(r) \simeq \frac{1}{2}m_pv^2 = 782 ~{\rm keV} \left(\frac{r}{3r_{\rm s}} \right)^{-1} \left( \frac{f_k}{0.1} \right)^2.
\label{Eq:ErHeavyPBH}
\end{equation}
\textcolor{black}{We take the velocity fraction of the outgoing proton wind to be $f_k = 0.2$}. 
 
The energy distribution function of the outflow is given by \cite{Takhistov2022JCAP}
\begin{align}
f (E) = & 3.54\times 10^{22} {\rm \frac{erg}{eV^2 ~sec}} s(2.35\times 10^8)^s \left(\frac{M_{\rm PBH}}{M_\odot} \right)^2 \nonumber \\
& \times \left( \frac{n_{\rm H}}{1~{\rm cm^{-3}}} \right) \left(\frac{\tilde v}{10~{\rm km/s}} \right)^{-3} \left( \frac{r_{\rm out}}{r_{\rm s}} \right)^{-s} f_v^{2s}E^{-s-1},
\label{Eq:fE_HeavyPBH}
\end{align}
where the duty cycle factor is set to 1, $\tilde v = (v^2 + c_{\rm s}^2)^{1/2} \simeq 10$ km/s in Leo T (with $v$ and $c_{\rm s} \simeq 9$ km/s representing the PBH velocity relative to the gas and the temperature-dependent sound speed in the gas, respectively), and $f_v$ denotes the velocity distribution of the PBHs.\textcolor{black}{\footnote{\textcolor{black}{Since the nature of the kinetic energy $E \propto v^2$ and the Keplerian velocity $v^2 \propto 1/r$, $E \rightarrow 0 $ corresponds to $r \rightarrow \infty$. In this study, $r_{\rm max} = r_{\rm out} = 100 r_{\rm s} \ll \infty$ and the energy distribution function $f(E) \propto 1/E^{s+1}$ cannot be divergent.}}} \textcolor{black}{Because} the lack of significant relative bulk velocity in Leo T, we assume a standard Maxwell-Boltzmann distribution in polar velocity coordinates as
\begin{equation}
\frac{df_v}{dv} = \sqrt{\frac{2}{\pi}}\frac{v^2}{\sigma_v^3}\exp\left(-\frac{v^2}{2\sigma_v^2}\right),
\label{Eq:MaxwellBoltzmann}
\end{equation}
where $\sigma_v$ is the velocity dispersion. Since the velocity dispersion of the PBHs is expected to match that of the gas, we take $\sigma_v = \sigma_g = 6.9$ km/s, giving an average velocity $\bar{v} = \sigma_v \sqrt{8/\pi} \simeq 11.0$ km/s. The parameter $s$ has been estimated by simulations to range from 0.4 \textcolor{black}{to} 0.8. In this study, we use $s=0.7$.  

\textcolor{black}{We note that, since the radius $r$ in Eq. (\ref{Eq:ErHeavyPBH}) is assumed to be $r=r_{\rm in}=3 r_{\rm s}$ or $r=r_{\rm out}=100 r_{\rm s}$, $E(r)$ is independent of the Schwarzschild radius $r_{\rm s}$. In a similar way, Eq.(\ref{Eq:fE_HeavyPBH}) is also independent of $r_{\rm s}$. }
\begin{figure}[t]
\centering
\includegraphics[keepaspectratio, scale=0.5]{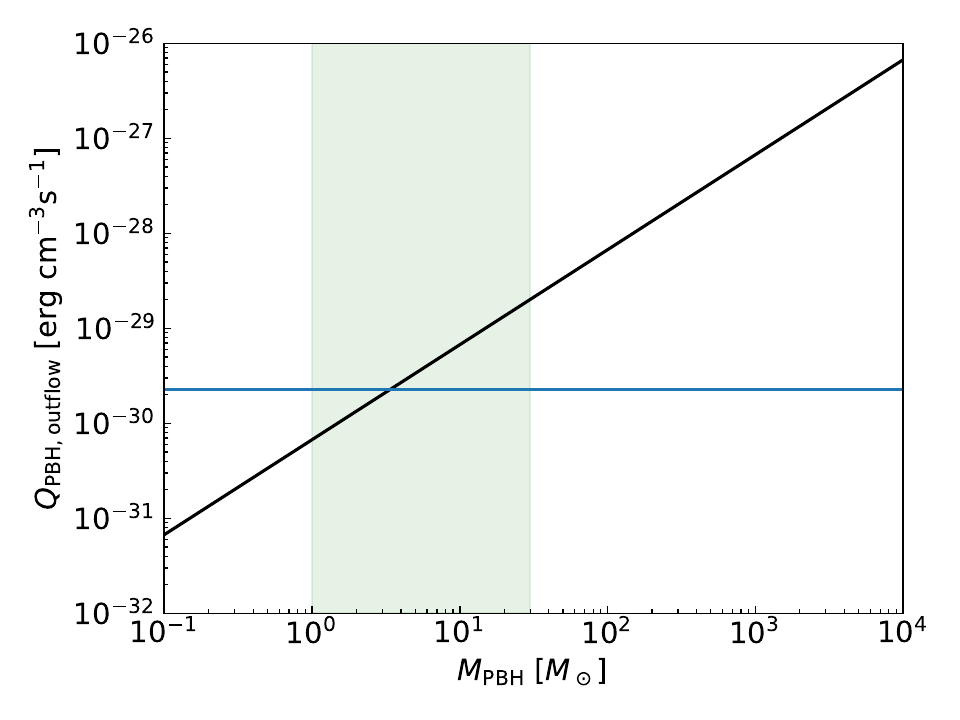}
 \caption{Gas heating rate in Leo T due to accretion mass outflow from heavy PBHs. The curve represents the heating rate under the assumption that all DMs consists of heavy PBHs. The horizontal line indicates the gas cooling rate.}
 \label{FIG:pbh_Q_acc_Msolar}
 \end{figure}

The heating rate per unit volume per unit time due to mass outflow from the accretion disk around a heavy PBH is given by
\begin{equation}
Q_{\rm PBH,outflow} = n_{\rm PBH} H_{\rm PBH,outflow}. 
\end{equation}

FIG. \ref{FIG:pbh_Q_acc_Msolar} illustrates the gas heating rate in Leo T due to accretion mass outflow from heavy PBHs. The horizontal line represents the gas cooling rate. As with FIG. \ref{FIG:pbh_Q_HR_gram}, the curve depicts the heating rate assuming all DMs consists of heavy PBHs, thereby showing the maximum possible gas heating rate under the specific case where $f_{\rm PBH} = 1$. 

It is important to note that heavy PBHs cannot constitute all of the DM. The maximum fraction of heavy PBH to DM is $f_{\rm PBH} \simeq 0.08$, which is only allowed within the narrow mass range of $1 \le M_{\rm PBH}/M_\odot \le 30$ in green band in FIG. \ref{FIG:pbh_Q_acc_Msolar}. In other mass regions, heavy PBHs can only make up a very small fraction of the DM \cite{Carr2021RPP,Takhistov2022JCAP}. Therefore, similar to the case with light PBHs, the actual heating ratio in Leo T must be adjusted downward from what is shown in FIG. \ref{FIG:pbh_Q_acc_Msolar}. However, unlike the light PBHs scenario, the case of heavy PBHs may yield significant results for the remainder of this paper, because $Q_{\rm PBH, outflow}$ can be comparable to $C$ in this mass range.

The actual maximum heating rate in light green band may be reduced by approximately 8\%. The expected ratio of heating to cooling is $0.27 \lesssim Q/C \lesssim 8.1$. Therefore, we can anticipate that gas heating due to heavy PBHs in Leo T could still provide substantial constraints, even when accounting for the current limits of $f_{\rm PBH}$. 

\subsection{Gas heating by dark photons \label{section:dark_photon}}
\begin{figure}[t]
\centering
\includegraphics[keepaspectratio, scale=0.5]{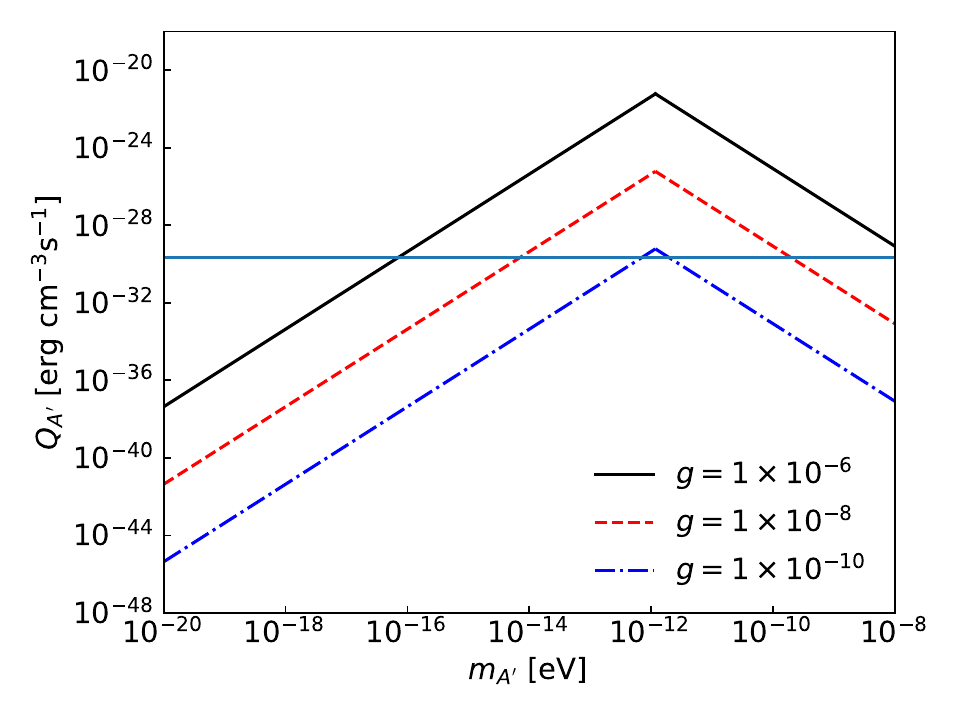}
 \caption{Gas heating rate in Leo T due to interactions with dark photons. The curve shows the heating rate for the specific case where all DMs consists of dark photons. The horizontal line represents the gas cooling rate.}
 \label{FIG:DP_Q}
 \end{figure}

In this subsection, we consider dark photons as the sole heating source for the gas in Leo T. Dark photons $A'$ are kinematically mixed with the standard model photon $A$. \textcolor{black}{The relevant Lagrangian for this study is \cite{Holdom1986PLB}
\begin{align}
\mathcal{L} = -\frac{1}{4}{F}^{\mu\nu}F_{\mu\nu} -\frac{1}{4}{F'}^{\mu\nu}F'_{\mu\nu} + \frac{m_{A'}^2}{2}A'^{\mu}A'_\mu 
 - \frac{g}{2}{F}^{\mu\nu}F'_{\mu\nu} + J_{\rm em}^\mu A_\mu,
\end{align}
where $m_{A'}$ denotes the mass of the dark photon, $g$ is the coupling parameter that describes the strength of the kinetic mixing, and $A_\mu$ ($A'_\mu$) and $F_{\mu\nu}$ ($F'_{\mu\nu}$) represent the gauge fields and field strengths of the visible (dark) photons, respectively. The electric current of the standard model is denoted by $J^\mu_{\rm em}$.}

\textcolor{black}{By redefining the gauge fields $A_\mu$ and $A'_\mu$, we can remove the kinetic mixing term ${F}^{\mu\nu}F'_{\mu\nu}$  \cite{Arias2012JCAP, Dubovsky2015JCAP}. In these bases, the kinetic mixing translates into} a direct coupling between the dark photon and the electric current. \textcolor{black}{According to the formalism in Refs. \cite{Dubovsky2015JCAP,Wadekar2021PRD}, the resulting Lagrangian is}
\begin{align}
\mathcal{L} = -\frac{1}{4}{F}^{\mu\nu}F_{\mu\nu} -\frac{1}{4}{F'}^{\mu\nu}F'_{\mu\nu} + \frac{m_{A'}^2}{2}A'^{\mu}A'_\mu 
 - \frac{e}{(1+g^2)^{1/2}}J^\mu_{\rm em} \left( A_\mu + gA'_\mu \right).
\end{align}

Dark photons generate an oscillating electric field that accelerates free electrons. These accelerated electrons then collide with ions, leading to heat dissipation in the plasma. The ionized region of Leo T behaves as a nonrelativistic plasma with a specific plasma frequency. The frequency of electrion-ion collisions is given by
\begin{equation}
\nu = \frac{4\sqrt{2\pi}\alpha_{\rm em}^2n_e}{3 m_e^{1/2}T_e^{3/2}} \ln \Lambda_{\rm C},
\end{equation}
where $n_e$ denote the number density of electron, \textcolor{black}{$T_e=6000$ K} is the electron temperature, \textcolor{black}{it is assumed to be virialized as in Eq.(\ref{Eq:MaxwellBoltzmann}) which corresponds to $\sigma$ = 6.9 km/s}, $\alpha_{\rm em} = 1/137$ is the fine structure constant, and $\ln \Lambda_{\rm C} = 0.5 \ln [4\pi T_e^3/ (\alpha_{\rm em}^3n_e)]$ is the Coulomb logarithm, respectively.

The potential energy of dark photon is converted into the kinetic energy of charged particles through electron-ion collisions. The resulting heating rate of the gas in Leo T per unit volume is given by
\begin{equation}
Q_{A'} = \nu \frac{g^2}{1+g^2} \rho_{A'}\times 
\begin{cases}
\left(\frac{m_{A'}}{\omega_{\rm p}}\right)^2 & {\rm for} \ m_{A'} \ll \omega_{\rm p} \\
\left(\frac{\omega_{\rm p}}{m_{A'}}\right)^2 & {\rm for} \ m_{A'} \gg \omega_{\rm p} \\
\end{cases},
\label{Eq:QA}
\end{equation}
where 
\begin{equation}
\omega_{\rm p} = \sqrt{\frac{4\pi \alpha_{\rm em}n_e}{m_e}},
\end{equation}
denotes the plasma frequency. 

FIG. \ref{FIG:DP_Q} illustrates the gas heating rate in Leo T resulting from interactions with dark photons. The curve depicts the heating rate under the assumption that all DMs is composed of dark photons. The horizontal line indicates the gas cooling rate.

\subsection{Gas heating by millicharged particles \label{section:Millicharged_particle}}
In this subsection, we consider millicharged particles as the sole heating source for the gas in Leo T. In the millicharged DM model, a millicharged particle $\chi$ possesses a small electric charge $\epsilon e$ where $e$ is the electron charge \cite{Holdom1986PLB}. There are two primary ways in which millicharged particles interact with the gas in Leo T \cite{Wadekar2021PRD}.

The first \textcolor{black} {way} is the Coulomb-like \textcolor{black}{interaction}, involving the millicharged particles and free electrons or ions. In the gas of Leo T, almost all ions are ionized hydrogen atoms ${\rm H^+}$ (proton $p$). The \textcolor{black}{total} Rutherford scattering cross section for a millicharged particle $\chi$ interacting with an electron or ionized hydrogen  is given by \cite{Wadekar2021PRD}
\begin{equation}
\sigma_{i\chi} = \frac{4\pi Z_i^2 \alpha_{\rm em}^2 \epsilon^2}{\mu_i^2 v_{\rm rel}^4} \ln (2\mu_i v_{\rm rel} \lambda_D) \quad {\rm for} \ i=(e^-, H^+),
\end{equation}
where $Z_i$ denotes the charge of the interacting particle,  $\mu_i = m_i m_\chi / (m_i + m_\chi)$ is the reduced mass, and $v_{\rm rel}$ represents the relative velocity between the millicharged particle and the gas component $i$. The ions are screened at distances greater than the Debye length, given by
\begin{equation}
\lambda_D = \sqrt{\frac{\textcolor{black}{T_e}}{4\pi \alpha_{\rm em} \sum_i Z_i^2 n_i}}.
\end{equation}

The second \textcolor{black}{way to be available} when the de Broglie wavelength of millicharged particle becomes smaller than the atom's screening length. In this regime, the millicharged particle can interact with the nucleus of neutral atoms, as the nucleus is screened over a distance approximately equal to the Bohr radius $a_0$. For hydrogen atoms, this interaction becomes significant when $m_\chi \gtrsim 0.07$ GeV.
In Leo T, neutral hydrogen and helium atoms are more prevalent than electrons or ions. Thus, for $m_\chi \gtrsim 0.07$ GeV, the dominant heat exchange by milicharged particles comes from neutral H and He atoms \cite{Wadekar2021PRD}. According to Refs. \cite{Kouvris2014PRD, Wadekar2021PRD}, the \textcolor{black}{total} cross section for a millicharged particle $\chi$ interacting  with hydrogen atom or helium atom is given by
\begin{align}
\sigma_{i\chi} = & \frac{2\pi Z_i^2 \alpha_{\rm em}^2 \epsilon^2}{\mu_i^2 v_{\rm rel}^4} \left[ \ln (1+4\mu_i^2 v_{\rm rel}^2 a^2) -\frac{1}{1+(4\mu_i^2 v_{\rm rel}^2a^2)^{-1} } \right] \quad  {\rm for} \ i=({\rm H, He}),
\label{Eq:sigmaMillicharged}
\end{align}
where $a=0.8853 a_0 Z_i^{-1/3}$ denotes the Thomas-Fermi radius. 

\begin{figure}[t]
\centering
\includegraphics[keepaspectratio, scale=0.5]{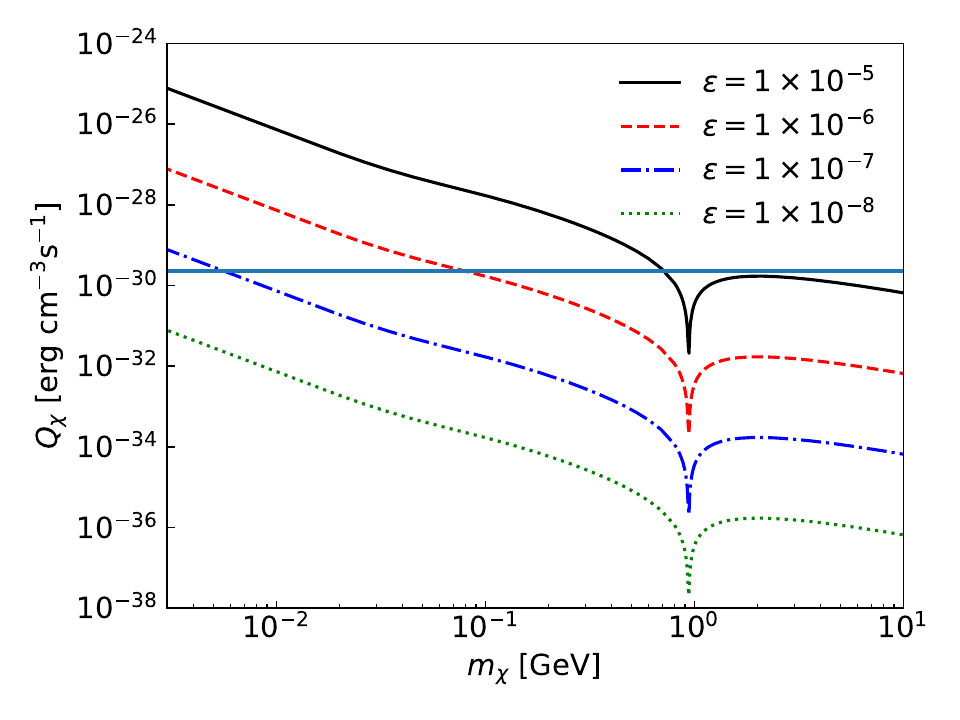} 
 \caption{Gas heating rate at Leo T due to interactions with millicharged particles. The curve represents the heating rate in the scenario where all DMs consists of millicharged particles. The horizontal line indicates the gas cooling rate.}
 \label{FIG:MC_Q}
 \end{figure}

We assume that both the millicharged particle $\chi$ and the gas component $i$ follow Maxwell-Boltzmann distributions in Leo T. Under this assumption, the millicharged particle can heat (or cool) gas if $T_i \neq T_\chi$. Since there is no relative bulk velocity between the millicharged particle and gas in Leo T, the heating rate of the gas component $i$ per unit volume per unit time can be expressed as \cite{Wadekar2021PRD,Dvorkin2014PRD}
\begin{equation}
Q_{i\chi} = \frac{2^{\frac{n+5}{2}} \Gamma(3+\frac{n}{2})}{\sqrt{\pi}(m_i+m_\chi)^2} \rho_i \rho_\chi  \sigma_{i\chi}^{(0)}  (T_\chi - T_i) \left(\frac{T_\chi}{m_\chi} - \frac{T_i}{m_i}\right)^{\frac{n+1}{2}},
\end{equation}
\textcolor{black}{where $\sigma_{i\chi}^{(0)} = \sigma_{i\chi} / v_{\rm rel}^{n}$. According to $\sigma_{i\chi} \propto v^{-4}_{\rm rel}$ which is shown in Eq. (\ref{Eq:sigmaMillicharged}), we use  $n=-4$.} Since the rms line-of-sight velocity dispersions of stars and the ${\rm H_I}$ gas are similar in Leo T, the velocity dispersions of the gas and the millicharged particle are approximately equal $v_\chi \simeq v_{\rm H_I}$. Consequently, thermal exchange between the millicharged particle and the gas depends on the mass differences because the millicharged particle and the proton, since $T_\chi - T_{\rm H_I} \simeq (m_\chi - m_p)v_{\rm H_I}^2$. Thus, the heating rate will be small for $m_\chi \simeq m_p \simeq 1$ GeV.

The total heating rate due to millicharged particles is given by 
\begin{equation}
Q_\chi = Q_{e^-\chi} + Q_{p^+\chi} + Q_{{\rm H}\chi} + Q_{{\rm He} \chi}.
\end{equation}
FIG. \ref{FIG:MC_Q} illustrates this heating rate, assuming all DMs are composed of millicharged particles. The curve represents the heating rate under this assumption, while the horizontal line indicates the gas cooling rate.

\section{Gas heating by both PBHs and DM particles \label{section:PBH_and_DM}}
\subsection{PBHs and dark photons \label{section:PBH_DP}}
Now, we consider a DM composition involving both PBHs and dark photons. In addition, we consider both PBHs and dark photons contribute to heating the interstellar medium gas in Leo T simultaneously. As discussed previously, the heating effect from light PBHs is negligible, so we focus on the contributions from heavy PBHs and dark photons. In this case, we have 4 model parameters to consider: the PBH fraction $f_{\rm PBH}$, the mass of PBH $M_{\rm PBH}$, the mass of dark photon $m_{A'}$, and the kinematic mixing parameter of dark photon $g$. Using the gas cooling rate and the heating rate for Leo T:
\begin{equation}
C \le Q_{\rm PBH, outflow}+Q_{A'},
\end{equation}
we explore the phenomenological constraints on these parameters.

First, we examine the relationship between the mass of the PBH $M_{\rm PBH}$ and the maximum allowed fraction of PBH $f_{\rm PBH}$.
In this study, we are interested in the scenario where both heavy PBHs and dark photons significantly contribute to heating the gas in Leo T.  Given that the heating rate due to dark photons is proportional to $Q_{A'} \propto (g m_{A'})^2$ for $m_{A'} \ll \omega_p$, see Eq.(\ref{Eq:QA}), and $g\ll 1$, the gas heating from dark photons might be comparable to the cooling rate ($C \simeq Q_{A'}$) in the manner of $g \propto m_{A'}^{-1}$ (for $m_{A'} \gg \omega_p$, $g \propto m_{A'}$). For instance, FIG. \ref{FIG:DP_Q} demonstrates that the combinations of $(g, m_{A'} [{\rm eV}]) \simeq (10^{-6}, 7\times 10^{-17}), (10^{-8}, 7\times 10^{-15})$ and $(10^{-10}, 7\times 10^{-13})$ result in $C \simeq Q_{A'}$. In this study, we use $g=10^{-8}$ and $m_{A'} \simeq 7 \times 10^{-15}$ eV as the  benchmark combination.

\begin{figure}[t]
\centering
\includegraphics[keepaspectratio, scale=0.5]{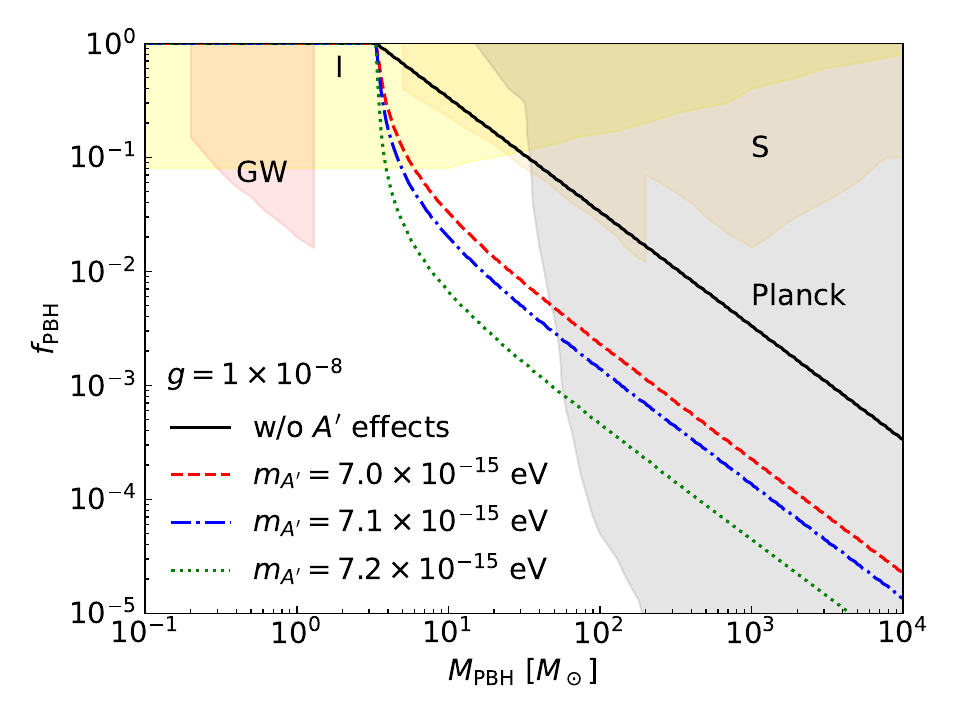} 
\includegraphics[keepaspectratio, scale=0.5]{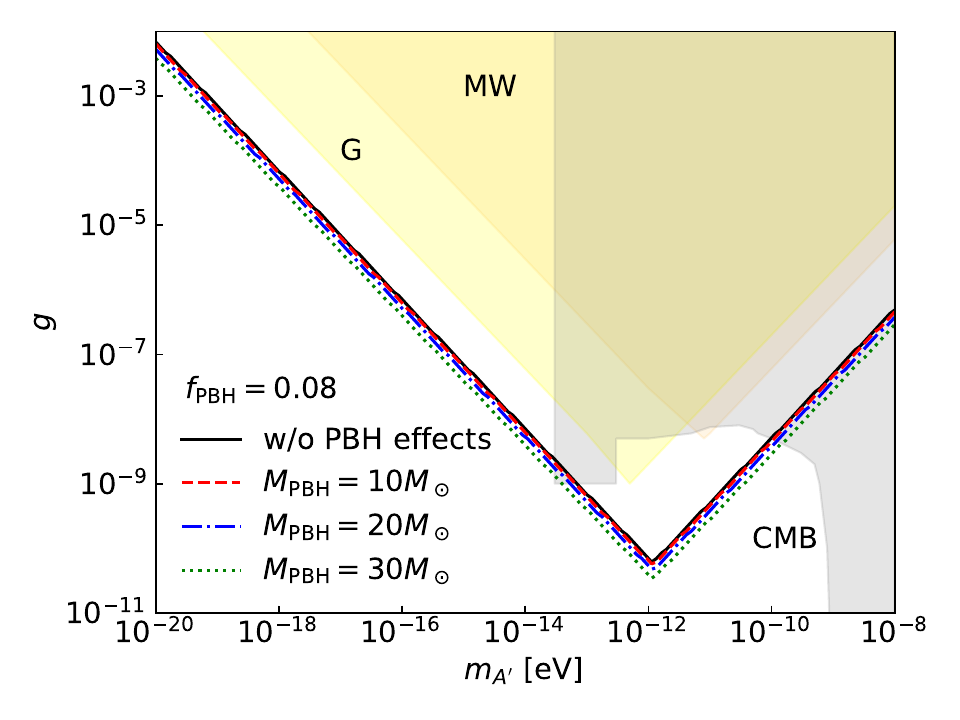} 
 \caption{Constraints on PBHs and dark photons from the Leo T dwarf galaxy. Top panel: upper limit on the PBH fraction $f_{\rm PBH}$. Bottom panel: upper limit on the kinetic mixing parameter $g$.}
 \label{FIG:pbh_DP}
 \end{figure}

The top panel of FIG. \ref{FIG:pbh_DP} illustrates the constraints on the PBH fraction $f_{\rm PBH}$ imposed by interstellar medium gas heating from both PBHs and dark photons in the Leo T dwarf galaxy. The black curve represents the upper limit of $f_{\rm PBH}$ when ignoring the gas heating effect of dark photons, as reported in Ref.\cite{Takhistov2022JCAP}. The remaining three curves show the upper limit of $f_{\rm PBH}$ considering gas heating from both PBHs and dark photons. The red, blue, and green curves correspond to dark photon masses $m_{A'}$ of  $7 \times 10^{-15}$ eV, $7.1 \times 10^{-15}$ eV, and  $7.2 \times 10^{-15}$ eV, respectively. The excluded regions for $f_{\rm PBH}$ are indicated from gravitational waves \cite{LIGOVIRGO2019PRL} (GW : red),  Icarus \cite{Oguri2018PRD} (I : yellow), combined bounds from surviving astrophysical systems in Eridanus II \cite{Brandt2016APJL}, Segue 1 \cite{Koushiappas2017PRL}, and disruption of wide binaries \cite{Rodriguez2014APJ} (S : orange), and Plank \cite{Haimoud2017PRD,Serpico2020PRR} (P : gray).

We find that the constraint on the PBH fraction are more restrictive for $4 \lesssim M_{\rm PBH}/M_\odot \lesssim 10^2$ compared to previously obtained constraints. This more stringent limit on $f_{\rm PBH}$ arises when the heating of the gas by dark photons is more effective than by PBHs. If dark photons are discovered in future experiments and their mass and the magnitude of the kinetic mixing parameter are determined, the relic abundance of PBHs in this mass range will be strongly constrained.

Next, we investigate the correlation between the upper bound on the mixing parameter $g$ and the mass of the dark photon $m_{A'}$. We focus on the scenario where both PBHs and dark photons contribute effectively to the gas heating in Leo T. We consider PBH masses in the range $1 \le M_{\rm PBH}/M_\odot \le 30$. In this mass range, the gas heating rate may be comparable to the cooling rate $C \simeq Q_{\rm PBH, outflow}$, see FIG. \ref{FIG:pbh_Q_acc_Msolar}. Additionally, we use $f_{\rm PBH} = 0.08$ as a benchmark case \cite{Carr2021RPP,Takhistov2022JCAP}. 

The bottom panel of FIG. \ref{FIG:pbh_DP} shows the constraints on the kinetic mixing parameter $g$ from interstellar medium gas heating by PBHs and dark photons in the Leo T dwarf galaxy. The black curve represents the upper limit on $g$ when ignoring the gas heating effects of dark photons, as previously discussed in Ref. \cite{Wadekar2021PRD}. The remaining three curves depict the upper limit on $g$ considering gas heating by both PBHs and dark photons. The red, blue, and green curves correspond to PBH masses of $10M_\odot$, $20M_\odot$, and $30M_\odot$, respectively. Excluded regions for $g$ are also shown, including constraints from the Milky Way interstellar medium \cite{Dubovsky2015JCAP} (MW : orange), the gas cloud G33.4-8.0 \cite{Wadekar2021PRD} (G : yellow), and the CMB \cite{Arias2012JCAP,Jaeckel2010ARNPS} (CMB : gray).

We observe that most significant constraint on the kinetic mixing parameter is already obtained even when ignoring the gas heating effect of PBHs \cite{Wadekar2021PRD}. When considering the gas heating effect of PBHs in Leo T, this constraint is modified but remains  the most significant.

\subsection{PBHs and millicharged particles \label{section:PBH_MC}}
In this subsection, we consider a DM model comprising heavy PBHs and millicharged particles. In this scenario, there are 4 model parameters: the PBH fraction $f_{\rm PBH}$, the mass of PBH $M_{\rm PBH}$, the mass of the millicharged particle $m_\chi$, and the charge parameter of the millicharged particle $\epsilon$. We examine phenomenological constraints on these parameters by analyzing the requirement for the gas cooling rate and heating rate in the Leo T:
\begin{equation}
C \le Q_{\rm PBH, outflow}+Q_\chi.
\end{equation}

First, we examine the constraint on the PBH fraction $f_{\rm PBH}$ in relation to the mass of the PBH $M_{\rm PBH}$. Similar to the case with PBHs and dark photons, we are interested in scenarios where both heavy PBHs and millicharged particles effectively contribute to the gas heating in Leo T. As illustrated in FIG. \ref{FIG:MC_Q}, combinations such as $(\epsilon, m_\chi [{\rm GeV}]) \simeq (10^{-5}, 0.7), (10^{-6}, 0.08)$ and $(10^{-7}, 0.005)$ result in $C \simeq Q_\chi$. For this study, we use $\epsilon=10^{-6}$ and $m_\chi \simeq 0.08$ as the benchmark combination. 

\begin{figure}[t]
\centering
\includegraphics[keepaspectratio, scale=0.5]{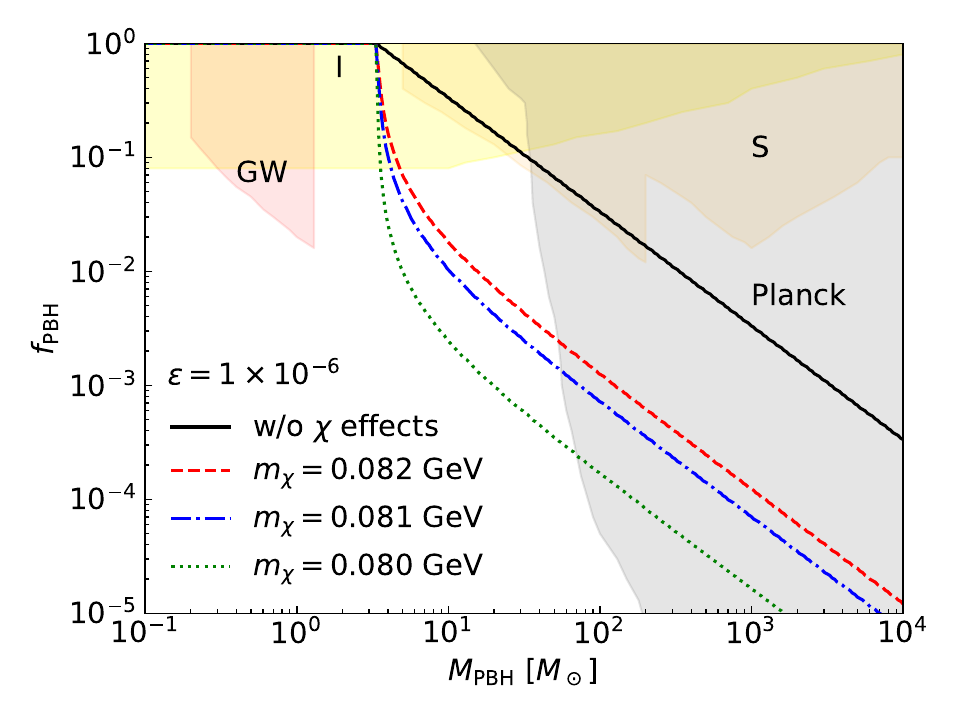} 
\includegraphics[keepaspectratio, scale=0.5]{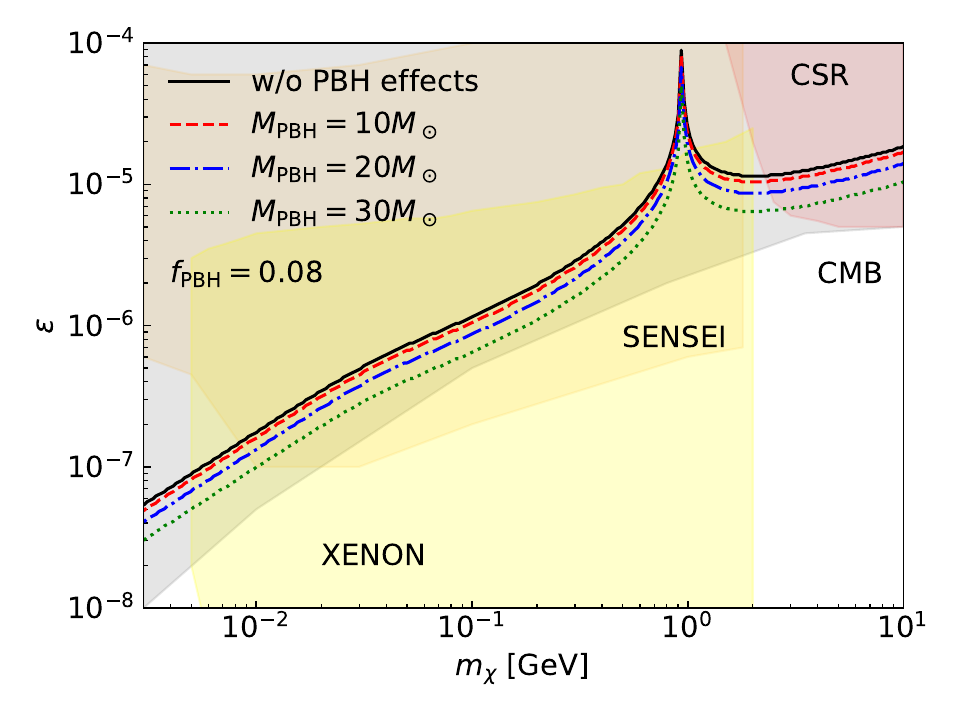} 
 \caption{Constraints on PBHs and millicharged particles from the Leo T dwarf galaxy. Top panel: upper limit on the PBH fraction $f_{\rm PBH}$. Bottom panel: upper limit on the charge parameter $\epsilon$.}
 \label{FIG:pbh_MC}
 \end{figure}

The top panel of FIG. \ref{FIG:pbh_MC} shows the constraints on the PBH fraction $f_{\rm PBH}$ from interstellar medium gas heating by PBHs and millicharged particles in the Leo T dwarf galaxy. The black curve represents the upper limit of the PBH fraction $f_{\rm PBH}$ when ignoring the gas heating effects of millicharged particles, as previously reported in Ref.\cite{Takhistov2022JCAP}. The remaining three curves illustrate the upper limit of $f_{\rm PBH}$ accounting for gas heating by both PBHs and millicharged particles. The red, blue, and green curves correspond to millicharged particle masses $m_\chi$ of $0.082$ GeV, $0.081$ GeV, and $0.80$ GeV, respectively. The excluded regions for $f_{\rm PBH}$ are indicated in the top panel of FIG. \ref{FIG:pbh_MC} as same as in FIG. \ref{FIG:pbh_DP}.

At first glance, it appears that constraint on the PBH fraction is more restrictive for $4 \lesssim M_{\rm PBH}/M_\odot \lesssim 10^2$ compared to previously obtained constraints. However, this initial impression is misleading. 

To clarify this, we examine the correlation between the upper bound on the charge parameter $\epsilon$ and the mass of the millicharged particle $m_\chi$. Since we are interested in cases where both PBHs and millicharged particles effectively contribute to the gas heating in Leo T, we consider $1 \le M_{\rm PBH}/M_\odot \le 30$ and $f_{\rm PBH} = 0.08$ as our benchmark case \cite{Carr2021RPP,Takhistov2022JCAP}. 

The bottom panel of FIG. \ref{FIG:pbh_MC} displays the constraints on the charge parameter $\epsilon$ from interstellar medium gas heating by PBHs and millicharged particles in the Leo T dwarf galaxy. The black curve represents the upper limit of the charge parameter $\epsilon$ when ignoring the gas heating effect of PBHs, as previously reported in Ref. \cite{Wadekar2021PRD}. The remaining three curves show the upper limits of $\epsilon$ when considering the gas heating contributions from both PBHs and millicharged particles. The red, blue, and green curves correspond to $M_{\rm PBH}$ = $10M_\odot$, $20M_\odot$, and $30M_\odot$, respectively. Excluded regions for $\epsilon$ are also indicated: SENSEI \cite{SENSEI2018PRL} (SENSEI : orange), XENON \cite{Essig2012PRL,Essig2017PRD} (XENON : yellow), CSR \cite{Mahdawi2018JCAP} (CSR : red), and CMB \cite{Wadekar2021PRD} (CMB : gray).

We observe that the masses of millicharged particles with charge parameters $\epsilon = 1\times 10^{-6}$ in the top panel of FIG. \ref{FIG:pbh_MC} are already excluded by SENSEI, CMB, and XENON constraints as shown in the bottom panel of FIG. \ref{FIG:pbh_MC}. From the bottom panel of FIG. \ref{FIG:pbh_MC}, the allowed maximum charge parameter $\epsilon \lesssim \mathcal{O}(10^{-8})$ for $m_\chi \lesssim 1$ GeV is very small, resulting in a low heating rate as depicted in FIG. \ref{FIG:MC_Q}. Consequently, it is challenging to derive more stringent constraints on the PBH fraction using the gas heating effects from millicharged particles in Leo T.

\section{Summary\label{section:summary}}
Given that many early universe models suggest the presence of PBHs, it is reasonable to assume that at least a portion of DM consists of PBHs. We consider a model where DM is composed of two components: PBHs and DM particles.

Recently, the Leo T dwarf galaxy has been utilized to examine the heating of interstellar medium gas by PBHs and DM particles. To ensure consistency with astrophysical observations, the heating rate of the gas due to these components must be equal to or less than the astrophysical cooling rate. Otherwise, the gas temperature would rise, potentially altering the ionization ratio. By analyzing the gas heating effect in Leo T, we can derive constraints on the PBHs as well as DM particles 

Previously studies have typically considered only one type of DM species as the source of heating for the interstellar medium gas. In contrast, we assumed that both PBHs and DM particles contribute simultaneously to gas heating. Specifically, we consider DM particles to be either dark photons or millicharged particles. We have demonstrated that heating effects due to Hawking radiation from light PBHs are negligible. Therefore, we focus exclusively on the heating mechanism associated with the accretion disk of heavy PBHs.

In the scenario where DM consists of both PBHs and dark photons, we have identified more restrictive constraint on the PBH fraction $f_{\rm PBH}$ for $4 \lesssim M_{\rm PBH}/M_\odot \lesssim 10^2$ compared to previously obtained limits. These constraints on $f_{\rm PBH}$ arise when the heating of the gas by dark photons is more significant than by PBHs. If dark photons are discovered and their mass and kinetic mixing parameter are precisely measured, the relic abundance of PBHs in this mass region will be strongly constrained. In contrast, when DM consists of PBHs and millicharged particles, we find that obtaining more stringent constraints on the PBH fraction is challenging due to the very small allowed values of the charge parameters in the model.

Finally, we would like to comment on the heating of the gas by free-free absorption \cite{Cyr2024MNRAS}. In a model that can produce a large number of low-frequency photons, these photons will be rapidly absorbed by the surrounding gas, even if the medium is only very slightly ionized. Hawking radiation from PBHs may produce many soft photons in some limits. In addition, conversions of dark photons into CMB photons (or vice versa) are more active at low frequencies, and can also produce many soft photons. These effects may give rise to the additional heating contribution, which may be important to consider. In this study, we neglect any soft photon heating. A detailed analysis of this issue is nedded in the future.

\vspace{3mm}







\end{document}